\documentstyle[12pt]{article}

\newcommand{\be}{\begin{equation}}
\newcommand{\ee}{\end{equation}}
\newcommand{\bea}{\begin{eqnarray}}
\newcommand{\eea}{\end{eqnarray}}

\newcommand{\p}[1]{(\ref{#1})}

\def\C{{\mathbf  C}}

\def\A{{\cal A}}
\def\G{{\cal G}}

\def\L{{\cal L}}

\begin{document}

\thispagestyle{empty}

\bibliographystyle{plain}
\begin{flushright}ENSLAPP-L-668/97\\
solv-int/yymmxxx\\
October 1997
\end{flushright}
\vskip 1.0truecm
\begin{center}{\bf\Large Supersymmetric Drinfeld-Sokolov reduction }
\end{center}
 \vskip 1.0truecm
\centerline{\bf F. Delduc, L. Gallot}
\vskip 1.0truecm
\begin{center} \it Laboratoire de Physique Th\'eorique ENSLAPP
\begin{footnote}{URA 14-36 du CNRS, associ\'ee  \`a l'ENS de Lyon et au LAPP\\
\hspace*{.6cm}Groupe de Lyon: ENS Lyon, 46 All\'ee d'Italie, 69364 Lyon, France
}\end{footnote}\end{center}
\vskip 2.0truecm  \nopagebreak

\begin{abstract}
The Drinfeld-Sokolov construction of integrable hierarchies, as well as its generalizations, may be extended to the case of loop superalgebras. A sufficient condition on the algebraic data for the resulting hierarchy to be invariant under supersymmetry transformation is given. The method used is a construction of the hierarchies in superspace, where supersymmetry is manifest. Several examples are discussed.
\end{abstract}
\setcounter{page}{0}
\newpage
\renewcommand{\thefootnote}{\arabic{footnote})}
\setcounter{footnote}{0}
\newpage

\section{Introduction}

One can find in the litterature many examples of supersymmetric hierarchies
(see for instance \cite{Mathieu1,math2,ina1,inami,morosi,kriso,krisoto}) 
but a general discussion of these hierarchies in the Drinfeld-Sokolov reduction framework is still lacking. The construction of a hierarchy in the Drinfeld-Sokolov approach relies on some algebraic data \cite{DS,Prin1,Prin2,DFG}. In particular, one needs that a loop algebra is introduced, together with an integer grading 
of this algebra, and a semi-simple positively graded element $\Lambda$ in the loop algebra. One may simply replace everywhere the words ``loop algebra'' by 
``loop superalgebra'', and the construction carries through without major difficulties. The hierarchy which will be obtained in this way  contains both commuting and anticommuting (Grassmann odd) fields. However,  in general it will not be supersymmetric, that is to say  it will not contain an odd conserved charge whose Poisson bracket with itself is the charge asociated with translation invariance. Another way to put it is that it will not admit a symmetry transformation relating  
commuting to anticommuting fields, in such a way that the commutator of two such transformations is a space translation. The simplest example of such a hierarchy will be given in section 3 \cite{kup1,gur}. 

Suppose now that we wish to construct supersymmetric hierarchies. It turns out that a very simple additional requirement on the input data is sufficient to ensure that the hierarchy will possess supersymmetry. Namely, the semi-simple element $\Lambda$,
necessarily an even element of the superalgebra, should be the square of an odd graded element $\Psi$ of the superalgebra. This will be shown in the next section, 
by constructing the hierarchy from the input data directly in superspace, where supersymmetry is manifest at all steps of the procedure. 

Several examples of the general construction are gathered in section 3. In particular, we shall give the Drinfeld-Sokolov formulations for the supersymmetric hierarchies studied in the Gelfand-Dickey approach in 
\cite{DG}. Some examples of supersymmetric extensions of the homogeneous hierarchies, which require the use of a twisted loop superalgebra,  are also studied.
 
\section{DS reduction}

In this section, we shall recapitulate the Drinfeld-Sokolov construction for the supersymmetric KdV type hierarchies in N=1 superspace. As a generalization of the construction in \cite{inami}, we shall give the supersymmetric version of the construction given in \cite{DFG}. Some details, common to both the bosonic and the supersymmetric cases will be omitted and we refer to the literature \cite{DS}, \cite{Prin1}, \cite{Prin2}, \cite{DFG} for details.

\paragraph{Superspace}

We shall consider an N=1 superspace with one space coordinate $x$ and one Grassmann coordinate $\theta$. The pair of coordinates $(x,\theta)$ will be denoted by $\tilde{x}$. The supersymmetric covariant derivative is defined by
\begin{equation}
D={\partial \over \partial\theta} + \theta\partial \, , \quad D^2 =\partial\equiv{\partial\over\partial x}.
\end{equation}
We define the integration of a superfield $H(\tilde{x})$ over the N=1 superspace to be
\begin{equation}
\int d\tilde{x} H(\tilde{x}) = \int dx DH\vert_{\theta=0}.
\end{equation}

A superalgebra ${\cal G}$ \cite{Kac} is a $Z_2$ graded algebra ${\cal G}={\cal G}_{\overline{0}}\oplus{\cal G}_{\overline{1}}$ with the product law $[,\}$ satisfying for any $a \in {\cal G}_{\alpha}$ and $b \in {\cal G}_{\beta}$
\begin{equation}
[a,b\} = -(-)^{\alpha\beta}[b,a\}; \qquad
[a,[b,c\}\}=[[a,b\},c\}+(-)^{\alpha\beta}[b,[a,c\}\}.
\end{equation}
We shall denote by hat the automorphism of ${\cal G}$ that reverses the sign of odd elements,
\begin{equation}
M\in {\cal G},\qquad M= M_{\overline{0}}+M_{\overline{1}} \rightarrow \hat{M}= M_{\overline{0}}-M_{\overline{1}}.
\end{equation}
We shall only consider classical superalgebras \cite{Kac} equipped with a bilinear invariant form denoted by $\langle , \rangle$.

In the following, we shall use superfields taking values in the tensor 
product $\underline{\cal G}$ of
a superalgebra ${\cal G}$ with some Grassmann algebra ${\cal G}r={\cal G}r_{\overline{0}}\oplus{\cal G}r_{\overline{1}}$. An element in this space is called even if it belongs to ${}_+{\cal G}={\cal G}_{\overline{0}}\otimes{\cal G}r_{\overline{0}}\oplus{\cal G}_{\overline{1}}\otimes{\cal G}r_{\overline{1}}$ and odd if it belongs to ${}_-{\cal G}={\cal G}_{\overline{0}}\otimes{\cal G}r_{\overline{1}}\oplus{\cal G}_{\overline{1}}\otimes{\cal G}r_{\overline{0}}$.
If we denote by $T^a$ a basis vector of $\cal G$, an element $A$ in $\underline{\cal G}$ may be expanded as
\begin{equation}
A=\sum_a A_a T^a,\qquad A_a\in {\cal G}r.
\end{equation} The supercommutator of two elements $A$ and $B$ in $\underline{\cal G}$ will be defined as
\begin{equation} [A,B\}=\sum_{a,b}A_aB_b[T^a,T^b\}.\end{equation}
With this definition, the supercommutator satisfies the symmetry properties listed in table 1.
\begin{center}
\begin{tabular}{|c|c|c|}\hline
$[A,B\}=$ & $A\in {}_+{\cal G}$ & $A\in {}_-{\cal G}$ \\ \hline
$B\in {}_+{\cal G}$ & $-[B,A\}$ & $-[\hat B,A\}$ \\ \hline
$B\in {}_-{\cal G}$ & $-[B,\hat A\}$ & $[\hat B,\hat A\}$\\ \hline
\end{tabular}\\
\vspace*{0.2cm}
Table 1
\end{center}
If $A(\tilde x)$ is a superfield taking values in ${}_+{\cal G}$ or ${}_-{\cal G}$, we extend the definition of the supercommutator in order to include the derivatives
\begin{eqnarray} &
[D,A(\tilde x)\}=\sum_a  \left( DA_a(\tilde x)\right) T^a, &\nonumber\\&
[D,A(\tilde x)\}=\mp[\hat A(\tilde x),D\}\quad \mbox{if}\quad A\in{}_\pm{\cal G}.&\nonumber\end{eqnarray}

We denote by ${\cal F}[J]\in {\cal G}r$ a functional of the superfield $J(\tilde{x})$ which takes values in ${}_+{\cal G}$ or ${}_-{\cal G}$. The functional derivative of ${\cal F}[J]$ with respect to $J(\tilde{x})$, denoted by ${\delta {\cal F}\over \delta J(\tilde{x})}$, is defined by 
\begin{equation}
{d \over d\epsilon}{\cal F}[J+\epsilon r]\vert_{\epsilon=0} = \int d\tilde{x}\langle {\delta {\cal F}\over \delta J(\tilde{x})}, r(\tilde{x})\rangle .
\end{equation}
where $r(\tilde{x})$ is any superfield with the same parity as $J(\tilde{x})$, and $\epsilon$ is a real parameter. Notice that the functional derivative of ${\cal F}$ has the same parity as ${\cal F}$ if $J$ is odd, and the opposite one if $J$ is even.

\paragraph{}
Following the lines of \cite{DFG}, a supersymmetric KdV system and its modified system will be associated with the choice of a sextuplet $({\cal A},d_1,d_0,\Psi;\alpha^0,\beta^0)$. ${\cal A}$ is an affine superalgebra
\begin{footnote}{We shall not restrict ourselves to affine superalgebras possessing a purely odd root system.}
\end{footnote} with vanishing center, that is  
a twisted loop superalgebra 
\be
\A=\L(\G,\tau)\subset \G\otimes \C [\lambda,\lambda^{-1}]
\label{2.1}\ee
 attached to a finite dimensional classical superalgebra ${\cal G}$
with an automorphism $\tau$ of finite order. $d_1$ and $d_0$ are two compatible integer gradations of $\A$ ($[d_1,d_0]=0$). The subsets ${\cal A}^{n}_{m}={\cal A}^{n}\cap{\cal A}_{m}$ with ${\cal A}^{n}= \{ X\in {\cal A}\vert d_1(X)=nX \}$, ${\cal A}_{m}= \{ X\in {\cal A}\vert d_0(X)=mX \}$ define a bi-grading of ${\cal A}$. We also assume that 
\begin{equation}
{\cal A}^{>0}\subset{\cal A}_{\geq 0}, \quad {\cal A}^{<0}\subset{\cal A}_{\leq 0}.\label{1.1.0}
\end{equation}

In the sextuplet, $\Psi$ is supposed to be an odd element of ${\cal A}$ with positive $d_1$-grade, $d_1(\Psi)=k\Psi$ with $k >0$, whose square $\Lambda ={1\over 2} [\Psi,\Psi\}$ is semisimple,
\begin{equation}
{\cal A} = \mbox{Ker}(\mbox{ad}\Lambda)\oplus \mbox{Im}(\mbox{ad}\Lambda).
\label{1.1.1}
\end{equation}
We denote by ${\cal Z}$ the center of the kernel ${\cal K}=\mbox{Ker}(\mbox{ad}\Lambda)$. ${\cal Z}$ contains only even elements of ${\cal A}$.

We finally assume that ${\cal A}^{0}$ can be decomposed into the direct sum of two of its subalgebras as
\begin{equation}
{\cal A}^{0}=\alpha^{0}\oplus\beta^{0}, \quad \mbox{with} \quad \alpha^0={\cal A}^{0}_{\geq 0}, \quad \beta^{0} \subset {\cal Z}^{0}.
\label{1.1.3}
\end{equation}
The superalgebra ${\cal A}$ is then decomposed into ${\cal A}= \alpha \oplus \beta$, where $\alpha={\cal A}^{>0}\oplus\alpha^{0}$ and $\beta=\beta^{0}\oplus{\cal A}^{<0}$ are subalgebras. The usual decomposition \cite{inami} is recovered if one requires the supplementary condition 
\be{\cal A}^{0}\subset{\cal A}_{0}, \label{strong}\ee
 which implies $\alpha^0={\cal A}^{0}$, $\beta^{0}=\{0\}$.

We introduce the  odd Lax operator 
\begin{equation}
{\cal L} = D +q(\tilde{x}) + \Psi,
\label{1.1.5}
\end{equation}
where the odd superfield $q(\tilde{x})$ takes values in ${}_-\A$.
We shall write evolution equations  on $\L$ of the zero curvature type
\begin{equation}
{\partial \over \partial t} {\cal L} = [A,{\cal L}\},
\label{1.1.7}
\end{equation}
where the even superfield $A(\tilde{x})$ takes values in ${}_+\A$.
From the odd Lax operator $\cal L$ we may obtain an even operator by
\begin{equation}
{\cal L}_x ={1\over 2}[\hat{\cal L},{\cal L}\}=\partial + Q -\Lambda
\label{1.1.9}
\end{equation}
where the even superfield $Q(\tilde{x})=Dq+{1\over 2}[\hat q,q\}-[\psi,q\}$ takes values in ${}_+\A$. It will evolve as
\begin{equation}
{\partial \over \partial t} {\cal L}_x = [\hat{A}, {\cal L}_x\}.
\label{1.1.11}
\end{equation}
The phase space
\begin{equation}
\Theta = \{ {\cal L} = D +q + \Psi \vert q \in {}_-{\cal A}^{<k}\cap{}_-{\cal A}^{\geq 0}\cap{}_-{\cal A}_{\geq 0} \}
\label{1.1.13}
\end{equation}
will define a modified KdV type system, while a KdV type system will be obtained from 
\begin{equation}
{\cal Q} = \{ {\cal L} = D +q + \Psi \vert q \in {}_-{\cal A}^{<k}\cap{}_-{\cal A}_{\geq 0} \}.
\label{1.1.15}
\end{equation}

In both cases, the construction relies on the following version of the formal dressing procedure. For any $q \in {}_-{\cal A}^{<k}$, there exists a unique $F(q(\tilde{x}))\in {}_+(\mbox{Im}(\mbox{ad}\Lambda))^{<0}$ such that
\begin{equation}
{\cal L}_{0} = e^{\hat{F}}{\cal L}e^{-F}= D + H +\Psi
\label{1.1.17}
\end{equation}
where $H\in{}_-(\mbox{Ker}(\mbox{ad}\Lambda))^{<k}$. At any finite grade, the components of both $F$ and $H$ are differential polynomials in terms of those of $q$. By this we mean that the graded components of $F$ and $H$ are polynomials in the components of $q$ and of a finite number of its derivatives $Dq$, $\partial q$, etc. Suppose that this is true up to grade $-n+k$ for $H$ and $-n$ for $F$, then, at the next grade, equation (\ref{1.1.17}) reduces to 
\begin{equation}
H^{k-n-1} = P(H^{k-1}, \cdots ,H^{k-n}, F^{-1},\cdots ,F^{-n}) + [\hat{F}^{-n-1},\Psi\}.
\label{1.1.19}
\end{equation}
It is not hard to show that the transformation $\mbox{ad}\Psi$ may
be restricted to the image $\mbox{Im}(\mbox{ad}\Lambda)$, and that this restriction is an isomorphism. In other words, there is a unique element 
$G^{-n-1-k}$ in ${}_+\mbox{Im}(\mbox{ad}\Lambda)$ such that
$$F^{-n-1}= [\hat{G}^{-n-1-k},\Psi\}.$$
Then equation \p{1.1.19} may be written as
\begin{equation}
H^{k-n-1} = P(H^{k-1}, \cdots ,H^{k-n}, F^{-1},\cdots ,F^{-n}) + [\Lambda,G^{-n-1-k}].
\label{1.1.21}
\end{equation}
From the fact that $\Lambda$ is semisimple, see eq.(\ref{1.1.1}), we conclude that there exists one and only one choice for $G^{-n-1-k}\in \mbox{Im}(\mbox{ad}\Lambda)$ such that this equation holds at this grade. For later purpose, we define for any constant $b\in {\cal Z}^{\geq 0}$ the functional
\begin{equation}
{\cal H}_b(q) = \int d\tilde{x}\langle b,H(q) \rangle.
\label{1.1.23}
\end{equation}
and the even superfield
\begin{equation}
B_b(q) = e^{-\hat{F}}be^{\hat{F}}.
\label{1.1.25}
\end{equation}
The graded components of $B_b(q)$ are differential polynomials in the components of $q$. Moreover, we have the property 
\begin{equation}
[b,{\cal L}_0]=0\quad\Rightarrow\quad [B_b(q),{\cal L}\}=0.
\label{1.1.27}
\end{equation}
According to a standard calculation, the functional derivative of ${\cal H}_b(q)$ is
\begin{equation}
{\delta \over \delta q}{\cal H}_b(q) = B_b(q).
\label{1.1.29}
\end{equation}

\paragraph{The modified system}

The formal dressing procedure may in particular be applied to the phase space $\Theta$ defined in eq. \p{1.1.13}. We shall construct on this phase space an integrable system of the modified KdV type. The splitting of ${\cal A}$ \begin{equation}
{\cal A} = \alpha +\beta
\end{equation}
yields an $r$-matrix 
\begin{equation}
{\cal R}= {1 \over 2}({\cal P}_{\alpha}-{\cal P}_{\beta})
\label{1.1.31}
\end{equation}
where ${\cal P}_{\alpha}$, ${\cal P}_{\beta}$ are the projectors on the respective subalgebras $\alpha$, $\beta$ along the subalgebras $\beta$, $\alpha$. ${\cal R}$ satifies the (super) classical modified Yang-Baxter equation
\begin{equation}
{\cal R}([X,{\cal R}(Y)\}+[{\cal R}(X),Y\})= [{\cal R}(X),{\cal R}(Y)\}+{1 \over 4}[X,Y\},
\label{1.1.33}
\end{equation}
which is a sufficient condition for the $r$-bracket $[ ,\}_{\cal R}$ defined by
\begin{equation}
[ X ,Y\}_{\cal R} =  [X,{\cal R}(Y)\}+[{\cal R}(X),Y\},
\label{1.1.35}
\end{equation}
to satisfy the super Jacobi identity. The evolution equation for the modified system, associated with the constant element $b \in {\cal Z}^{\geq 0}$, is given by
\begin{equation}
{\partial \over \partial t_b}q = [{\cal R}(B_b(q)),{\cal L}\}
.\label{1.1.37}
\end{equation}
using eq.(\ref{1.1.27}), this evolution equation may be written as
\begin{equation}
{\partial \over \partial t_b}q = [ {\cal P}_{\alpha}(B_b(q)),{\cal L}\}=  -[{\cal P}_{\beta}(B_b(q)),{\cal L}\} ,
\end{equation}
from which one easily checks that 
(\ref{1.1.37}) is a consistent evolution equation. Using that $[B_a(\theta), B_b(\theta)]=0$ for all $a,b\in {\cal Z}^{\geq 0}$
and that, as a 
consequence of (\ref{1.1.37}),  
${\partial \over \partial t_a} B_b(\theta)=[{\cal R}(B_a(\theta)), B_b(\theta)\}$, 
together with the modified classical Yang-Baxter equation (\ref{1.1.33})
for ${\cal R}$, it can be shown that the flows 
associated with different elements of ${\cal Z}^{\geq 0}$ commute:
\begin{equation}
\left[ {\partial \over \partial t_a}\,,\, {\partial \over \partial t_b} \right] q=0
\qquad \forall\,  a,b\in {\cal Z}^{\geq 0}.
\label{1.1.39}
\end{equation}
Finally, the functionals ${\cal H}_a(q)$ are conserved quantities for this evolution equation
\begin{equation}
{\partial \over \partial t_b}{\cal H}_a(q)=0.
\label{1.1.41}
\end{equation}
 
We shall now describe the hamiltonian formalism for this hierarchy of evolution equations. For this, it will be convenient to introduce a Poisson bracket on the local functionnals of the fields in the space 
\begin{equation}
{\cal M} = \{ {\cal L} = D +q + \Psi \vert q \in {}_-{\cal A}^{<k} \}.
\label{1.1.43}
\end{equation}
which contains as subsets both $\Theta$ and ${\cal Q}$. The Poisson bracket is inferred from the existence of the $r$-bracket $[,\}_{\cal R}$ and reads
\begin{eqnarray}
\{ f,g \}_{\cal R} &=& \int d\tilde{x}
\left\langle \left[ {\delta f \over \delta q}, {\delta g \over \delta q} \right\}_{\cal R}, q+\Psi \right\rangle  \nonumber \\
&+&
(-1)^{[g]+1}\left\langle {\cal R} {\delta f \over \delta q}, D\hat{{\delta g \over \delta q}} \right\rangle
+
(-1)^{[g]+1}\left\langle  {\delta f \over \delta q}, D{\cal R}\hat{{\delta g \over \delta q}} \right\rangle,
\label{1.1.45}
\end{eqnarray}
where $[g]=0$ if $g$ is an even functional, and $[g]=1$ if $g$ is odd. The important fact here is that $\Theta \subset {\cal M}$ is a Poisson submanifold, 
so that we may restrict the Poisson bracket to $\Theta$. The evolution equation (\ref{1.1.37}) may be written in the following hamiltonian form
\begin{equation}
{\partial \over \partial t_b}f = \{ f, {\cal H}_b(q) \}_{\cal R},
\label{1.1.47}
\end{equation}
and the hamiltonians are in involution with respect to the Poisson bracket
\begin{equation}
\{ {\cal H}_a(q),{\cal H}_b(q)\}_{\cal R}
=0.
\label{1.1.49} 
\end{equation}

\paragraph{The KdV type system}

We shall describe here the construction of a KdV type system and its relation with the modified system. We shall require the nondegeneracy condition
\begin{equation}
\mbox{Ker}(\mbox{ad}(\Lambda))\cap {\cal A}^{<0}_{0}=\{ 0 \}\label{1.1.51}.
\end{equation}
The phase space of the KdV type system is the factor space ${\cal Q}/{\cal N}$ where ${\cal N}$ is the group of gauge transformations $e^{\gamma}$ acting on ${\cal Q}$ as
\begin{equation}
e^{\gamma}: {\cal L}\mapsto e^{\hat{\gamma}}{\cal L}e^{-\gamma}, \quad \gamma(\tilde{x}) \in {}_+{\cal A}^{<0}_{0}.
\label{1.1.53}
\end{equation}
Let us define a subspace ${\cal Q}_V$ of the phase space ${\cal Q}$ by
\begin{equation}
{\cal Q}_V = \{ {\cal L} = D+q_V+\Psi \vert q_V(\tilde{x}) \in {}_-V \}
\label{1.1.55}
\end{equation}
where $V$ is a vector space such that
\begin{equation}
{\cal A}^{<k}\cap{\cal A}_{\geq 0} = [ \Psi , {\cal A}^{<0}_{0} \} \oplus V.
\label{1.1.57}
\end{equation}
Due to the nondegeneracy condition (\ref{1.1.51}), for any field configuration $q$ in ${\cal Q}$, there is a unique field-dependent gauge transformation such that the image is ${\cal Q}_V$. In other words, ${\cal Q}_V$ is a model for ${\cal Q}/{\cal N}$. When regarded as functions on ${\cal Q}$, the components of $q_V(q)$ are differential polynomials which freely generate the set of gauge invariant differential polynomials on ${\cal Q}$. We shall refer to ${\cal Q}_V$ as a Drinfeld-Sokolov gauge.

To construct an integrable hierarchy on ${\cal Q}/{\cal N}$, we shall first exhibit commuting flows on ${\cal Q}$ by means of the formal dressing procedure.
Similarly to the modified case, we define commuting flows on ${\cal Q}$ by
\begin{equation}
{\partial \over \partial t_b}q = [{\cal R}(B_b(q)),{\cal L}\}
,\label{1.1.59}
\end{equation}
for any $b \in {\cal Z}^{\geq 0}$. It is not a difficult task to verify that the conditions on the sextuplet 
$({\cal A},d_1,d_0,\Psi;\alpha^0,\beta^0)$ ensure that (\ref{1.1.59}) gives a consistent evolution equation on ${\cal Q}$. This equation has a gauge invariant meaning, which is to say that the evolution of gauge invariant quantities is given by gauge invariant differential polynomials. This is a consequence of the uniqueness of the formal dressing procedure \cite{DFG}. If we use ${\cal Q}_V$ as a model for ${\cal Q}/{\cal N}$ , then the evolution equation (\ref{1.1.59}) on ${\cal Q}_V$ takes the form
\begin{equation}
{\partial \over \partial t_b}q_V = [{\cal R}(B_b(q_V))+\eta_b(q_V),D+q_V+\Psi\}
,\label{1.1.61}
\end{equation}
where $\eta_b(q_V)\in {}_+{\cal A}^{<0}_{0}$ is a uniquely determined differential polynomial in $q_V$. The flows thus defined on ${\cal Q}_V$  commute. 

The hamiltonian interpretation of these flows is as follows. Notice first that ${\cal Q}\subset {\cal M}$ is a Poisson submanifold with respect to the Poisson bracket $\{ , \}_{\cal R}$, hence the evolution equation (\ref{1.1.59}) can be put in the hamiltonian form
\begin{equation}
{\partial \over \partial t_b}f = \{ f ,{\cal H}_b \}_{\cal R}
\label{1.1.63}
\end{equation}
where ${\cal H}_b$ is obtained from the formal dressing procedure. We know that if $f$ is gauge invariant then the right hand side of (\ref{1.1.63}) is gauge invariant. Since ${\cal H}_b$ is a gauge invariant functional of ${\cal Q}$, by the uniqueness property, we are then led to suspect that the Poisson bracket $\{ f,g\}_{\cal R}$ of any two gauge invariant functionals on ${\cal Q}$ is again gauge invariant. Close inspection of the Poisson bracket shows that this is indeed the case. As a consequence, a non-linear Poisson bracket is defined on ${\cal Q}/{\cal N}$ by identifying the local functionals on ${\cal Q}/{\cal N}$ with gauge invariant functionals on ${\cal Q}$. The KdV type hierarchy on ${\cal Q}/{\cal N}$ is generated by the hamiltonians ${\cal H}_b$ and this induced bracket. The evolution equations of this hierarchy (\ref{1.1.61}) hence take the form (\ref{1.1.63}) where $f$ is a gauge invariant functional. 

The modified type system is related to the KdV type system by a Miura map. One can first check that $\Theta \in {\cal Q}$ and ${\cal Q}/{\cal N}$ have the same dimension
\begin{equation}
\mbox{dim}\left( {\cal A}^{<k}\cap {\cal A}^{\geq 0}\cap{\cal A}_{\geq 0}\right) =\mbox{dim}\left( {\cal A}^{<k}\cap {\cal A}_{\geq 0}\right) -\mbox{dim}\left( {\cal A}^{< 0}\cap{\cal A}_0\right)
\label{1.1.65}
\end{equation}
which means that the two systems have the same number of independent fields. The map 
\begin{equation}
\mu : \Theta \rightarrow {\cal Q}/{\cal N}
\label{1.1.67}
\end{equation}
induced by the projection ${\cal Q}\rightarrow {\cal Q}/{\cal N}$ is the Miura map. $\mu$ maps the flows of the modified system onto those of the KdV type system. From the hamiltonian point of view, $\mu$ is a Poisson map from the linear Poisson bracket $\{ , \}_{\cal R}$ on $\Theta$ to the non-linear Poisson bracket on ${\cal Q}/{\cal N}$ induced by the Poisson bracket $\{ , \}_{\cal R}$ on ${\cal Q}$.

\noindent {\it Remark} In this construction, the hamiltonians are associated with the elements in ${\cal Z}$, the center of $\mbox{Ker}(\mbox{ad}\Lambda)$. Inami and Kanno \cite{inami} associated the hamiltonians with the elements in a vector space $\tilde{\cal K}$ such that the following decomposition of $\mbox{Ker}(\mbox{ad}\Lambda)$ holds 
\begin{equation}
{\cal K}= [ {\cal K},{\cal K} \} \oplus \tilde{\cal K}.
\end{equation}
$[ {\cal K},{\cal K} \}$ is the commutant of ${\cal K}$. Indeed, an evolution equation of the hierarchy may be written on the dressed operator ${\cal L}_{0}$ and reads  
\begin{equation}
\partial_{t_b}{\cal L}_{0}=[A_b, {\cal L}_{0}\},
\end{equation}
where $A_b \in {\cal K}$. If one decomposes the dressed operator into ${\cal L}_{0}=D+H_{[ {\cal K},{\cal K} \}}+H_{\tilde{\cal K}}+\Psi$, then one obtains the conservation law
\begin{equation}
\partial_{t_b}H_{\tilde{\cal K}}=-D(\hat{A}_b)_{\tilde{\cal K}}
\quad\Rightarrow\quad \partial_{t_b}\int d\tilde x H_{\tilde{\cal K}}=0.
\end{equation}
In the examples that we shall look at hereafter, the set of conserved quantities associated with $\tilde{\cal K}$ is identified with the one associated with ${\cal Z}$. This is a consequence of the fact that $\tilde{\cal K}$ may be choosen in a way such that ${\cal Z}$ and $\tilde{\cal K}$ are isotropic and conjugated with respect to the invariant bilinear form $\langle , \rangle$ on ${\cal A}$.

\section{Examples}

\subsection{A reminder of the scalar formalism}

In this section, we wish to study the matrix Lax formulation for the two series of N=2 KdV hierarchies described in \cite{DG}. Before starting, we recall some basic facts about the N=1 scalar Lax formulation of these hierarchies.

This formalism, which is an analog of the Gelfand-Dickey formalism, is based on the use of the algebra ${\cal C}$ of even N=1 pseudo differential operators ($\Psi$DOs) of the form
\begin{equation}
L= \sum_{i <M} u_{i}D^{i}.
\label{2.1.2}
\end{equation}
The coefficient functions $u_i$ are N=1 superfields. We define the residue of the pseudo-differential operator $L$ by 
$\mbox{res} L=u_{-1}$. The trace of $L$ is the integral of the residue
\begin{equation}
\mbox{Tr}L=\int \mbox{d}\tilde{x}\,\mbox{res}L, \quad
\mbox{Tr}[L,L']=0.
\label{2.1.4}
\end{equation}
The algebra ${\cal C}$ can be decomposed into the direct sum of two associative subalgebras ${\cal C} = {\cal C}_{\geq 1} \oplus {\cal C}_{\leq 0}$ where $L=L_{\geq 1}$ is in ${\cal C}_{\geq 1}$ if it is a differential operator containing only strictly positive powers of $D$, and $L=L_{\leq 0}$ is in ${\cal C}_{\leq 0}$ if it is a pseudo-differential operator containing only negative or zero powers of $D$. 
We denote by ${\cal R}$ the classical $r$-matrix associated with this splitting:
\begin{equation}
{\cal R}(L) = {1 \over 2}(L_{\geq 1}-L_{\leq 0}).
\label{2.1.10}
\end{equation}
Then the commuting flows of an N=2 KP hierarchy, for an operator $L$ of the form
\begin{equation}
L= D^{2n} +\sum_{i =1}^{\infty} U_i D^{2n-i-1},
\label{2.1.6}
\end{equation}
are defined by 
\begin{equation}
\partial_{t_k}L=[{\cal R}(L^{k\over n}),L].
\label{2.1.8}
\end{equation}
 
This N=2 KP hierarchy is the so-called non-standard supersymmetric KP hierarchy \cite{dasb1,ghosh}.
Due to the fact that the $r$-matrix ${\cal R}$ is not skew symmetric, these flows are hamiltonian with respect to a linear and two quadratic Poisson brackets. We introduce these Poisson structures on the linear functionals of $L$ of the type $l_X(L)=\mbox{Tr} ( L X)$, $X$ being some element in ${\cal C}$ which does not depend on the phase space fields $\{ U_i \}$. The linear structure is  
\begin{equation}
\{ l_X, l_Y \}_{(1)} (L)= \mbox{Tr} \left( L [X,Y]_{{\cal R}}\right),
\label{2.1.12}  
\end{equation}
and the two quadratic ones \cite{dasp,DG}are
\begin{eqnarray}
\{ l_X , l_Y \}_{(2)}^{a}(L)= \mathrm{Tr}(LX(LY)_{\geq 0}
-XL(YL)_{\geq 0})
+\int d^2\tilde{x} (-\psi_{Y} \mathrm{res}[L,X]\nonumber\\
+\mathrm{res}[L,Y]\,
 \mathrm{res}(XLD^{-1}) 
- \mathrm{res}[L ,X]\,
 \mathrm{res}(YLD^{-1})
),&
\label{2.1.14} \\
\{ l_X , l_Y \}_{(2)}^{b}(L)= \mathrm{Tr}(LX(LY)_{\geq 0}
-XL(YL)_{\geq 0})
+\int d^2\tilde{x} (\psi_Y \, \mathrm{res}[L,X] \nonumber\\
+\mathrm{res}[L ,Y]\,
 \mathrm{res}(LXD^{-1}) 
- \mathrm{res}[L ,X]\,
 \mathrm{res}(LYD^{-1})
)&\label{2.1.16}
\end{eqnarray}
where the quantity $\psi_{X}$ is defined up to a constant by
$D\psi_{X}= \mathrm{res}[L,X]$. The two quadratic brackets are of the $abcd$ type \cite{Maillet}. Introducing the hamiltonians
${\cal H}_{k} = {n\over k}\mbox{Tr}(L^{k\over n})$, the N=2 KP evolution equations (\ref{2.1.8})
may be written in the hamiltonian form
\begin{equation}
\partial_{t_k}l_X(L)=\{ l_{X}, {\cal H}_{k+n} \}_{(1)} (L)=\{ l_{X}, {\cal H}_{k}\}_{(2)}^{a,b} (L).
\label{2.1.18}
\end{equation}
The hamiltonians form a family of quantities in involution with respect to any of these three brackets.
The two quadratic brackets are related to each other by an invertible Poisson map $p$ of ${\cal C}$ defined by
\begin{equation}
p(L) =  \partial^{-1}L^{t},
\label{2.1.20}
\end{equation}
where $L^{t}$ is the operator adjoint to $L$.
One has the relation
\begin{equation}
\{ l_{X} \circ p,l_{Y}\circ p \}_{(2)}^{a} =-\{ l_{X},l_{Y} \}_{(2)}^{b}\circ p,
\label{2.1.22}
\end{equation}
which gives an equivalence between the two quadratic structures. However, there is no relation between the hamiltonians $\mbox{Tr}(L^{ k \over n})$ and $\mbox{Tr}(p(L)^{ k \over n-1})$.

As in the bosonic case, N=2 KdV hierarchies are consistent reductions of the KP hierarchy. Indeed, a first series of KdV hierarchies \cite{ina1,inami} is obtained by considering the restriction $L=L_{\geq 1}$, that is to say 
\begin{equation}
L=D^{2n} +\sum_{i =1}^{2n-2} w_iD^{2n-i-1},
\label{2.1.24}
\end{equation}
which defines a Poisson submanifold of the quadratic bracket $\{ ,\}_{(2)}^{a}$. The Poisson algebra on the fields $w_i$ is the N=2 ${\cal W}_{n}$ algebra.
The restriction 
\begin{equation}
L=D^{2n-2} +\sum_{i =1}^{2n-3} w_iD^{2n-i-3}+D^{-1}w_{2n-2}
\label{2.1.26}
\end{equation}
defines a Poisson submanifold of both the quadratic bracket $\{ ,\}_{(2)}^{b}$ and the linear one $\{ ,\}_{(1)}$ and then gives a second  series of KdV hierarchies. The two series of KdV hierarchies are related to each other by the Poisson map $p$. 

The two series of N=2 KdV hierarchies correspond to two series of N=2 modified KdV hierarchies as we shall now explain. 

Let us write first the operator (\ref{2.1.24}) in the factorized form
\begin{eqnarray}
L=&(D+\theta_{2n-2})(D+\theta_{2n-2}+\theta_{n-1})(D+\theta_{2n-3}+\theta_{n-1})\cdots \nonumber \\
 &(D+\theta_{n}+\theta_{2})(D+\theta_{n}+\theta_{1})(D+\theta_{1})D,
\label{2.1.36}
\end{eqnarray}
where the Miura fields $\theta_k$ satisfy the following Poisson algebra of the Gardner type
\begin{equation}
\{ f , g \} = -\int d\tilde{x}\sum_{i,j=1}^{2n-2}C_{ij}D\left( {\delta f \over \delta \theta_{i}}\right){\delta g \over \delta \theta_{j}}.
\label{2.1.38}
\end{equation}
The symmetric invertible matrix $C$ is 
\begin{equation}
C= \left( 
\begin{array}{cc}
 0 & A \\ 
A^t & 0
\end{array}
\right),
\end{equation}
where the $(n-1)\times (n-1)$ upper triangular matrix $A$ has ones on the diagonal and above the diagonal: $A_{ij}=0$ if $j>i$ and $A_{ij}=1$ if $j \leq i$.
The comparison between equations (\ref{2.1.24}) and (\ref{2.1.36}) yields the Miura transformation which gives the KdV fields in terms of the Miura fields. A standard calculation \cite{Dickey1} shows that the Miura transformation is a Poisson map from the Gardner type bracket (\ref{2.1.38}) to the first quadratic bracket $\{ , \}_{(2)}^{a}$. 

The similar statement for the second series of KdV hierarchies is the following.
Let us write the operator (\ref{2.1.26}) in the factorized form
\begin{eqnarray}
L=D^{-1}(D+\theta_{1})(D+\theta_{n}+\theta_{1})(D+\theta_{n}+\theta_{2})\cdots \nonumber \\
(D+\theta_{2n-3}+\theta_{n-1})(D+\theta_{2n-2}+\theta_{n-1})(D+\theta_{2n-2}) \label{2.1.40}
\end{eqnarray}
which yields the Miura transformation by comparison with (\ref{2.1.26}). The Miura fields satisfy the Gardner Poisson algebra (\ref{2.1.38}). Then, the Miura transformation is a (anti) Poisson map from the Gardner type bracket (\ref{2.1.38}) to the second quadratic bracket $\{ , \}_{(2)}^{b}$.

\subsection{A matrix formulation for the first series}

The first series of N=2 KdV hierarchies has been studied in \cite{ina1,inami}. It can be described within the usual Drinfeld-Sokolov formalism with an appropriate choice of the sextuplet $({\cal A},d_1,d_0,\Psi;\alpha^{0},\beta^{0})$ such that $\beta^{0}=\{ 0 \}$. 

We consider ${\cal A}= sl(n\vert n)\otimes \mathbf{C}[\lambda , \lambda^{-1}]$.
A $2n\times 2n$ matrix in $sl(n\vert n)$ has vanishing supertrace
$$ \mbox{str}M=\sum_{i=1}^{2n} (-1)^{i+1}M_{ii}. $$
The center of the superalgebra ${\cal A}$ is spanned by the matrices $\lambda^k I$ where $I$ is the $2n\times 2n$ identity matrix. ${\cal A}$ is endowed with the two compatible gradations
\begin{eqnarray}&
d_1 = 2n\lambda\partial_{\lambda}+ \mathrm{ad}{\cal K}_{I}, \quad {\cal K}_{I}= \sum_{k=1}^{2n}{2n+1-2k \over 2}e_{k,k}  &\label{2.2.2}\\&
d_0 = 2\lambda\partial_{\lambda}+ \mathrm{ad}{\cal K}, \quad {\cal K}= \sum_{k=1}^{2n-1}e_{k,k}. &\label{2.2.4}
\end{eqnarray}
Here $d_1$ is the principal gradation but $d_0$ is not the homogeneous gradation. Notice that this choice for $d_0$ is the only point where our formalism differs from that of \cite{ina1,inami}. We have made this choice in order to be able to define a Drinfeld-Sokolov gauge. The end result, though, is the same as in \cite{ina1,inami}. The assumptions in (\ref{1.1.0}) are satisfied, as well as the supplementary condition ${\cal A}^{0}\subset{\cal A}_{0}$. We choose for $\Psi$ the odd element of $d_1$-grade one
\begin{equation}
\Psi = \sum_{k=1}^{2n-1}e_{k,k+1}+\lambda e_{2n,1} \label{2.2.8}
\end{equation}
whose square $\Lambda$ is semisimple.

\paragraph{The modified system} We shall first identify the modified system defined by the above choice of algebraic data with the system characterized by the factorized Lax operator (\ref{2.1.36}). We parametrize the phase space $\Theta$ in (\ref{1.1.13}) by 
\begin{equation}
q(\tilde{x}) = \sum_{k=1}^{n-1} \theta_{k}(e_{2k-1,2k-1}+e_{2k,2k})+
\sum_{k=n}^{2n-2} \theta_{k}(e_{2k-2n+2,2k-2n+2}+e_{2k-2n+3,2k-2n+3}).\label{2.2.14}
\end{equation}
We have fixed the gauge freedom induced by the center of ${\cal A}$ by setting the coefficient of $e_{2n,2n}$ in $q$ to zero.
The explicit evaluation of the Poisson bracket (\ref{1.1.45}) on $\Theta$ yields exactly the bracket given in (\ref{2.1.38}). The linear problem ${\cal L}\Psi=0$, where $\Psi = (\psi_{1},\psi_2, \cdots \psi_{2n})^{t}$, leads to the eigenvalue equation 
\begin{eqnarray}
L\psi_{2n}=&(D+\theta_{2n-2})(D+\theta_{2n-2}+\theta_{n-1})(D+\theta_{2n-3}+\theta_{n-1})\cdots \nonumber \\
 &(D+\theta_{n}+\theta_{2})(D+\theta_{n}+\theta_{1})(D+\theta_{1})D\psi_{2n} 
=\lambda \psi_{2n}.\label{2.2.16}
\end{eqnarray}
This allows to identify the hamiltonians of both systems and finally the two modified systems themselves.

\paragraph{The KdV type system} It is easily checked that the non-degeneracy condition \p{1.1.51} is satisfied in this case at hand. In order to identify the KdV type system associated with the above algebraic data as the $n^{th}$ N=2 KdV hierarchy of the first series, we parametrize the phase space ${\cal Q}/{\cal N}$ by a convenient DS gauge ${\cal Q}_V$ whose general element is
\begin{equation}
q_V(\tilde{x})=\sum_{k=1}^{2n-2}(-)^{k}U_{k}e_{2n-1,k}.\label{2.2.18}
\end{equation}
From the linear problem ${\cal L}\Psi=0$, we obtain the eigenvalue equation on $\psi_{2n}$
\begin{equation}
L\psi_{2n}= (D^{2n}+U_1D^{2n-2}+U_2D^{2n-3}+\cdots  +U_{2n-2}D)\psi_{2n}=\lambda\psi_{2n}.\label{2.2.20}
\end{equation}
The correspondence between the two parametrizations of $L$ in (\ref{2.2.16}) and (\ref{2.2.20}) provides the Miura map $\mu: \Theta \rightarrow {\cal Q}/{\cal N}={\cal Q}_V$ which is a Poisson map when $\Theta$ is equipped with the linear bracket (\ref{1.1.45}) and ${\cal Q}/{\cal N}$ is equipped with the non-linear bracket obtained as the reduction of the linear bracket on ${\cal Q}$. Using the identification between the linear bracket on $\Theta$ with the Gardner type bracket in (\ref{2.1.38}), we conclude that the non-linear bracket on ${\cal Q}/{\cal N}$ coincide with the bracket $\{ ,\}_{(2)}^{a}$ on the operator L. The identification with the $n^{th}$ N=2 KdV hierarchy is then established.

\subsection{A matrix formulation for the second series}

The second series of N=2 KdV hierarchies can be recovered within the generalized  Drinfeld-Sokolov formalism with an appropriate choice of the sextuplet
$({\cal A},d_1,d_0,\Psi;\alpha^{0},\beta^{0})$.  

We consider ${\cal A}= sl(n\vert n-1)\otimes \mathbf{C}[\lambda , \lambda^{-1}]$.
An $2n-1\times 2n-1$ matrix in $sl(n\vert n-1)$ has vanishing supertrace
$$ \mbox{str}M=\sum_{i=1}^{2n-1} (-1)^{i+1}M_{ii}. $$
 ${\cal A}$ is endowed with the two compatible gradations
\begin{eqnarray}
&d_1 = (2n-2)\lambda\partial_{\lambda}+ \mathrm{ad}{\cal K}_{II}, \quad {\cal K}_{II}= \sum_{k=1}^{2n-1}(n-k)e_{k,k}  &\label{2.3.2}\\&
d_0 = \lambda\partial_{\lambda} &\label{2.3.4}
\end{eqnarray}
satisfying the assumptions in (\ref{1.1.0}), but not the suplementary condition \p{strong}. Here $d_0$ is the homogeneous gradation but $d_1$ is not the principal gradation. We choose for $\Psi$ the odd element of $d_1$-grade one
\begin{equation}
\Psi = \sum_{k=1}^{2n-2}e_{k,k+1}+\lambda (e_{2n-1,1} +e_{2n,2})\label{2.3.8}
\end{equation}
whose square $\Lambda$ is semisimple. The $d_1$-grade zero part of the center ${\cal Z}$ of the kernel is spanned by the matrix 
\begin{equation}
\Lambda_0=\lambda e_{2n-1,1}+ e_{1,1}+2 \sum_{k=2}^{2n-2} e_{k,k}+e_{2n-1,2n-1}+\lambda^{-1} e_{1,2n-1},
\label{2.3.14}
\end{equation}
and we take $\beta^{0}=\mathrm{span}\{ \Lambda_0 \}$, so that the decomposition
(\ref{1.1.3}) holds.

\paragraph{The modified system}
We wish to identify the modified system defined by the above choice of algebraic data with the system belonging to the factorized Lax operator (\ref{2.1.40}). For this we now parametrize the phase space $\Theta$ in (\ref{1.1.13}) by 
\begin{eqnarray}
&q(\tilde{x}) = \xi H +\sum_{k=1}^{n-1}\theta_k(e_{2n-2k,2n-2k}+e_{2n-2k+1,2n-2k+1}) &\nonumber\\
&+ \sum_{k=n}^{2n-2}\theta_k(e_{4n-3-2k,4n-3-2k}+e_{4n-2-2k,4n-2-2k})+\lambda (\theta_{1}+\theta_{2n-2})e_{2n-1,1},&\label{2.3.16}
\end{eqnarray}
where $H=(n-1)(\sum_{k=0}^{n-1}e_{2k+1,2k+1})+n(\sum_{k=1}^{n-1}e_{2k,2k})$ is an element in ${\cal K}^{0}$ which has the property that it cannot be written as the commutator of two elements in ${\cal K}$. $H$ is also dual to $\Lambda_0$ since $\langle H , \Lambda_0 \rangle = 2(1-n) \neq 0$. Let us define the linear functional of $q$
\begin{equation}
{\cal F}[q] \equiv \int \mathrm{d}\tilde{x}f(\tilde{x})\langle \Lambda_0 , q \rangle = 2(1-n)\int \mathrm{d}\tilde{x}f(\tilde{x}) \xi.
\end{equation}
Using the property that $\Lambda_0$ is an element of $\beta^{0}$ and the expression of the Poisson structure (\ref{1.1.45}), it is not difficult to show that the functional ${\cal F}[q]$ has vanishing Poisson bracket with any other functional ${\cal G}[q]$. Hence $\xi$ can be set to zero. The explicit evaluation of the Poisson bracket (\ref{1.1.45}) on $\Theta$ then yields exactly the bracket given in (\ref{2.1.38}). The linear problem ${\cal L}\Psi=0$  leads to the eigenvalue equation 
\begin{eqnarray}
& L\psi_{1}=D^{-1}(D+\theta_{1})(D+\theta_{1}+\theta_{n})
(D+\theta_{2}+\theta_{n})\cdots &\nonumber \\
&(D+\theta_{2n-3}+\theta_{n-1})(D+\theta_{2n-2}+\theta_{n-1})(D+\theta_{2n-2})\psi_1 =2\lambda \psi_{1}.&\label{2.3.18}
\end{eqnarray}
This allows to identify the hamiltonians of both systems and finally the two modified systems themselves.

\paragraph{The KdV type system} In order to identify the KdV type system associated with the above algebraic data as the $n^{th}$ N=2 KdV hierarchy of the second series, we parametrize the phase space ${\cal Q}/{\cal N}$ by a convenient DS gauge ${\cal Q}_V$ whose general element is
\begin{equation}
q_V(\tilde{x})=\sum_{k=1}^{2n-2}(-)^{k}U_{k}e_{2n-1,2n-1-k} +\xi H.\label{2.3.20}
\end{equation}
The field $\xi$ can be set to zero by identical considerations as in the modified case. From the linear problem ${\cal L}\Psi=0$, we obtain the eigenvalue equation on $\psi_{1}$
\begin{equation}
L\psi_{1}= D^{-1}(D^{2n-1}+U_1D^{2n-3}+\cdots U_{2n-3}D+U_{2n-2})\psi_{1}=2\lambda\psi_{1}.\label{2.3.22}
\end{equation}
The correspondence between the two parametrizations of $L$ in (\ref{2.3.18}) and (\ref{2.3.22}) provides the Miura map $\mu: \Theta \rightarrow {\cal Q}/{\cal N}={\cal Q}_V$. The identification with the $n^{th}$ N=2 KdV hierarchy is then established by the same arguments as in the case of the first series.

\section{Other examples} 
In this section we shall give the Drinfeld-Sokolov reduction procedure for some interesting, supersymmetric or not, integrable systems.

\subsection{Fermionic extension of KdV} 

The fermionic extension of KdV \cite{kup1,gur} is a non supersymmetric bi-hamilto\-nian integrable hierarchy. Its second hamiltonian structure is the N=1 superconformal algebra. The evolution equations of the hierarchy are
\begin{equation}
\partial_{t_k}L = [L^{2k+1\over n}_{\geq 0},L]
\end{equation}
where the Lax operator $L$ is
\begin{equation}
L = \partial^{2}-u-\xi\partial^{-1}\xi ,
\end{equation}
where $\xi (x)$ is a Grassmann odd field. This hierarchy can be recovered within the usual (non supersymmetric) Drinfeld-Sokolov formalism based on a superalgebra rather than a Lie algebra. We consider the loop superalgebra ${\cal A}=osp(1\vert 2) \otimes \mathbf{C}[\lambda,\lambda^{-1}]$. The supertrace of the $3\times 3$ matrix $M$ is defined by 
\begin{equation}
\mathrm{str}(M)=M_{1,1}-M_{2,2}+M_{3,3}
\end{equation}
and a matrix in $osp(1\vert 2)$ may be parametrized as
\begin{equation}
M =
\left( 
\begin{array}{ccc}
a & \alpha & b \\
\beta & 0 & \alpha \\
c & -\beta & -a
\end{array}
\right) .
\end{equation}
${\cal A}$ is equipped with the two compatible gradations
\begin{eqnarray}
& d_1 = 4\lambda \partial_{\lambda} +\mathrm{ad} \, (e_{11}-e_{33}) ,&\\
& d_0 = \lambda \partial_{\lambda}. &
\end{eqnarray}
$d_1$ is the principal gradation and $d_0$ is the homogeneous one. They satisfy the conditions \p{1.1.0} as well as \p{strong}. Finally, we choose for $\Lambda$ the even $d_1$-grade two element 
\begin{equation}
\Lambda = e_{1,3}+\lambda e_{3,1}.
\end{equation}
$\Lambda$ is semisimple. Notice that $\Lambda$ cannot be written as the square of an odd element in ${\cal A}$. This is the reason why the system associated with this choice of $\Lambda$ is not supersymmetric.
The lax operator is 
\begin{equation}
{\cal L} = \partial + Q + \Lambda 
\end{equation}
with $Q \in {\cal A}^{\leq 1}\cap {\cal A}_{\geq 0}$. The gauge algebra is the set of $\lambda$ independent lower triangular matrices and a convenient DS gauge is
\begin{equation}
Q_V= 
\left( 
\begin{array}{ccc}
0 & 0& 0 \\
\xi & 0 & 0 \\
u & -\xi & 0
\end{array}
\right) .
\end{equation}
The linear problem ${\cal L}\Psi=0$ where $\Psi = (\psi_1,\psi_2,\psi_3)^{t}$ leads to the eigenvalue equation
\begin{equation}
L \psi_1 = (\partial^{2}-u-\xi\partial^{-1}\xi)\psi_1=\lambda \psi_1 .
\end{equation}
The relation with the fermionic extension of KdV is then obtained by using standard arguments.

\subsection{N=4 KdV, Extensions of N=2 KdV}

\paragraph{N=4 KdV}

The N=4 KdV hierarchy \cite{delivan,lax4,DG} can be viewed as a reduction of the N=2 KP hierarchy \cite{DG} for which the scalar Lax operator in $N=1$ superspace is
\begin{equation}
L = \partial -W_1 +D^{-1}W_2 + D^{-1}\bar\phi D^{-1}\phi
\label{sn4}\end{equation}
where $W_1$, $\phi$ and $ \bar \phi$ are even superfields, while $W_2$ is an odd superfield. This hierarchy can be recovered within the generalized DS formalism. We consider the algebra ${\cal A}= sl(2\vert 2) \otimes \mathbf{C}[\lambda , \lambda^{-1}]$. The supertrace of a $4\times 4$ matrix $M$ is
$$ \mbox{str}M=M_{11}-M_{22}-M_{33}+M_{44}.$$ This superalgebra is equipped with the two compatible gradations
\begin{eqnarray}
& d_1 = 2\lambda \partial_{\lambda} +\mathrm{ad} \, \mathrm{Diag}(1,0,0,-1), &\\
& d_0 = \lambda \partial_{\lambda}. &
\end{eqnarray}
We choose the $d_1$-grade one odd element $\Psi$ to be
\begin{equation}
\Psi = e_{1,2}+e_{2,4}+\lambda e_{2,1}+\lambda e_{4,2}.
\end{equation}
$\Lambda=\Psi^2$ is semisimple. We then define 
\begin{equation}
\beta^{0}={\cal Z}^{0}=\mathrm{span}(\Lambda_0)  
\end{equation} 
where $ \Lambda_0 = \lambda e_{4,1} + e_{1,1}+2e_{2,2}+e_{4,4}+\lambda^{-1}e_{1,4}$. The decomposition (\ref{1.1.3}) holds.
A convenient DS gauge for this system is then
\begin{equation}
{\cal L}= D + 
\left( 
\begin{array}{cccc}
u + \xi & 1 & 0 & 0 \\
\lambda + W_1 & u - \xi& 0 & 1 \\
\phi & 0 & u -\xi & 0 \\
W_2 & \lambda & \bar\phi & u + \xi
\end{array}
\right) .
\end{equation}
The superfield $u$ is the coefficient of the center of $sl(2\vert 2)$, hence it can be set consistently to zero. The superfield $\xi$ is the coefficient of an element $H$ which belongs to ${\cal K}^{0}$, dual to $\Lambda_0$. Hence $\xi$ is central in the Poisson algebra and can be set consistently to zero. The linear problem ${\cal L}\Psi =0$ then leads to the eigenvalue equation
\begin{equation}
L \psi_1 = (\partial -W_1 +D^{-1}W_2 + D^{-1}\bar\phi D^{-1}\phi
)\psi_1 = 2 \lambda \psi_1.
\end{equation}
leading to the identification of this matrix hierarchy with the N=4 KdV hierarchy. One can find in \cite{DG} an infinite series of hierarchies where the $N=1$ scalar Lax operator has the same pseudo-differential part as in eq.\p{sn4}, but where the differential part is more general. A matrix Lax formulation for these hierarchies is found by generalizing the above scheme to the superalgebra $sl(n \vert n)$.

\paragraph{Fermionic extension of the N=2 a=4 KdV hierarchy}

The fermionic extension of the N=2 a=4 KdV hierarchy is very similar to the N=4 KdV hierarchy just discussed. It is a reduction of the N=2 KP hierarchy and its scalar lax operator \cite{DG} is 
\begin{equation}
L = \partial -W_1 +D^{-1}W_2 + D^{-1}\bar\phi D^{-1}\phi  
\end{equation}
where $W_1$ is an even superfield while $W_2$, $\phi$ and $\bar\phi$ are odd superfields.  

Its matrix interpretation is also very similar to the one of the N=4 KdV hierarchy as we shall see now. We choose the loop superalgebra ${\cal A}= sl(3\vert 1) \otimes \mathbf{C}[\lambda , \lambda^{-1}]$.  The supertrace of a $4\times 4$ matrix $M$ is
$$ \mbox{str}M=M_{11}-M_{22}+M_{33}+M_{44}.$$
 ${\cal A}$ is equipped with the two compatible gradations
\begin{eqnarray}
& d_1 = 2\lambda \partial_{\lambda} +\mathrm{ad} \, \mathrm{Diag}(1,0,0,-1) &\\
& d_0 = \lambda \partial_{\lambda}. &
\end{eqnarray}
We choose the $d_1$-grade one odd element $\Psi$ to be
\begin{equation}
\Psi = e_{1,2}+e_{2,4}+\lambda e_{2,1}+\lambda e_{4,2}.
\end{equation}
The even element $\Lambda=\Psi^2$ is semisimple. We then take 
\begin{equation}
\beta^{0}={\cal Z}^{0}=\mathrm{span}(\Lambda_0)  
\end{equation} 
where $ \Lambda_0 = \lambda e_{4,1} + e_{1,1}+2e_{2,2}+e_{4,4}+\lambda^{-1}e_{1,4}$. The decomposition (\ref{1.1.3}) holds.

A convenient DS gauge for this system is then
\begin{equation}
{\cal L}= D + 
\left( 
\begin{array}{cccc}
u +\xi & 1 & 0 & 0 \\
\lambda + W_1 & u +2\xi& 0 & 1 \\
\phi & 0 & -u & 0 \\
W_2 & \lambda & \bar\phi & u +\xi
\end{array}
\right). 
\end{equation}
The element $H=e_{1,1}+2e_{2,2}+e_{4,4}$ belongs to ${\cal K}^{0}$ and is dual to $\Lambda_0$. Hence $\xi$ is central in the Poisson algebra and can be set consistently to zero. The abelian gauge invariance generated by the grade zero element $K=e_{1,1}+e_{2,2}-e_{3,3}+e_{4,4}$ which commutes with $\Psi$ can be used to set $u$ to zero. The linear problem ${\cal L}\Psi=0$ then leads to the eigenvalue equation
\begin{equation}
L \psi_1 = ( \partial -W_1 +D^{-1}W_2 + D^{-1}\bar\phi D^{-1}\phi)\psi_1 = 2 \lambda \psi_1.
\end{equation}

This scheme can be generalized to $sl(n+1 \vert n-1)$ without difficulty, giving a matrix Lax formulation for some hierarchies described in \cite{DG}.

\subsection{N=2 homogeneous hierarchies}

Homogeneous bosonic hierarchies \cite{DS,Prin1,Prin2} are associated with a loop algebra ${\cal A}={\cal G}\otimes \mathbf{C}[\lambda,\lambda^{-1}]$ for which one takes $d_1=d_0$ to be the homogeneous gradation. As a consequence, in this case there exists only an analogue of the modified KdV type equations. The simplest such hierarchies have a Lax operator of the type 
\begin{equation}
{\cal L} = \partial + q + \Lambda
\end{equation}
where the semisimple grade one element $\Lambda=\lambda H$ is such that $H$ belongs to the Cartan subalgebra of ${\cal G}$. In this case $q$ takes its values in ${\cal A}_{0}={\cal G}$. The flows of the hierarchy take the particular form 
\begin{equation}
\partial_{t_b}q = [(B_b[q])_{-1},\lambda H]
\end{equation}
which means that only the components of $q$ in $(\mathrm{Im}(\mathrm{ad}\Lambda))_{0}$ have non trivial evolution equations. The components of $q$ in $(\mathrm{Ker}(\mathrm{ad}\Lambda)_{0})$ can be consistently set to zero at the level of evolution equations. If $H$ is regular, $(\mathrm{Ker}(\mathrm{ad}\Lambda)_{0})$ is the Cartan subalgebra of ${\cal G}$ but if $H$ is not regular, $(\mathrm{Ker}(\mathrm{ad}\Lambda)_{0})$ is a bigger subalgebra of ${\cal G}$.

The simplest examples are associated with ${\cal G}=sl(2)$ or $sl(3)$. When ${\cal G}=sl(2)$, the above construction leads to the non-linear Schr\"odinger (NLS) hierarchy with the Lax operator
\begin{equation}
{\cal L} = \partial + 
\left( 
\begin{array}{cc}
0 & \phi \\
\bar \phi & 0
\end{array}
\right) +  \lambda \left( 
\begin{array}{cc}
1 & 0\\
0 & -1
\end{array}
\right) .\label{Laxnls}
\end{equation}
When ${\cal G}$ is the rank two Lie algebra $sl(3)$, there exists a one parameter family of such integrable hierarchies. Each hierarchy is associated with a Cartan element 
\begin{equation}
H_t = \mathrm{Diag}(1,-2t,2t-1).
\end{equation} 
In the generic case, $H_t$ is regular. The phase space contains six fields associated with the roots of $sl(3)$. In the particular cases when $t = -{1 \over 2}, {1 \over 4}$ or $1$, $H_t$ is not regular and the phase space contains only four fields.

In the following, we shall study two examples of N=2 supersymmetric extensions of the homogeneous hierarchies. We consider the twisted loop superalgebra  
\begin{equation}
{\cal A} = \bigoplus_{k \in \mathbf{Z}} \lambda^{2k}{\cal G}_{\overline{0}}\oplus\lambda^{2k+1}{\cal G}_{\overline{1}},
\end{equation}
where ${\cal G}$ is a classical superalgebra, and $d_1=d_0$ is the homogeneous gradation. The choice ${\cal G}=sl(2 \vert 1)$ will lead to the N=2 NLS hierarchy. In the case when ${\cal G}=sl(3 \vert 1)$, there are only three possible choices of the odd element $\Psi$. They lead to generalizations of the $sl(3)$ hierarchies associated with the particular values $t=-{1 \over 2}, {1 \over 4}, 1$.

\paragraph{The N=2 NLS hierarchy} 
The N=2 NLS hierarchy \cite{roelo} can be viewed as a reduction of the N=2 KP hierarchy \cite{DG} associated with the N=1 scalar Lax operator 
\begin{equation}
L = \partial - \phi D^{-1} \bar\phi D  
\end{equation}
where $\phi$ and $\bar\phi$ are fermionic fields. It can be obtained in the  DS formalism as a homogeneous hierarchy for which ${\cal G}=sl(2 \vert 1)$. We choose the supertrace of a $3\times 3$ matrix $M$ to be
$$\mbox{str}M=M_{11}-M_{22}+M_{33}.$$
 We choose the grade one odd element $\Psi$ to be
\begin{equation}
\Psi = \lambda ( e_{1,2}+ e_{2,1}) .
\end{equation}
$\Lambda = \Psi^2 = \lambda^2(e_{1,1}+e_{2,2})$ is a semisimple element of ${\cal A}$. Notice that $e_{1,1}+e_{2,2}$ is not the Cartan element appearing in the bosonic operator (\ref{Laxnls}). It differs from it by ${1 \over 2}(e_{1,1}+2e_{2,2}+e_{3,3})$, which is an element of the center of ${\cal G}_{\overline{0}}$. Since $\Psi$ has grade one, the phase space belongs to the grade zero part of the superalgebra, so that the matrix Lax operator ${\cal L}$ contains only odd superfields
\begin{equation}
{\cal L}= D + 
\left( 
\begin{array}{ccc}
u_0 & \lambda & \phi \\
\lambda & u_0+u_1 & 0  \\
\bar\phi & 0 & u_1 
\end{array}
\right) .
\end{equation}
The grade zero element $H=e_{1,1}+e_{2,2}$ which belongs to the center of the kernel ${\cal Z}$ generates an abelian gauge invariance. A consequence of this gauge invariance is that the field $u_1$ has trivial equations of evolution
\begin{equation}
\partial_{t_b}u_1 = -\partial_{t_b}\mathrm{str}(Hq)=-\mathrm{str}(H[(B_b(q))_{-1},\Psi \}) =0
\end{equation}
for any $b$ in ${\cal Z}({\cal K})^{>0}$. Hence $u_1$ can be set consistently to zero in the evolution equations. The gauge invariance can also be used to set $u_0=0$. The linear problem ${\cal L}\Psi =0$ then leads to the eigenvalue equation
\begin{equation}
L \psi_2= ( \partial - \phi D^{-1} \bar\phi D ) \psi_2 = \lambda^2\psi_2 
\end{equation}
and the hierarchy thus obtained is the N=2 NLS hierarchy.

\paragraph{The $sl(3 \vert 1 )$ example}

We choose here ${\cal G}=sl(3 \vert 1 )$.
We choose the supertrace of a $4\times 4$ matrix $M$ to be
$$\mbox{str}M=M_{11}+M_{22}+M_{33}-M_{44}.$$
 The only three possible choices of the element $\Psi$ are 
\begin{equation}
\Psi_{i}= \lambda ( e_{i,4}+ e_{4,i}), \quad i=1,2,3.
\end{equation}
The semisimple \'element $\Lambda=\Psi^{2}$ is then
\begin{equation}
\Lambda_{i} = {\lambda^2  \over 3}( U +(3 \delta_{i,1}-1)H_{t_i} ) , \quad   i=1,2,3.
\end{equation}
where $U =\mathrm{Diag}(1,1,1,3)$ belongs to the center of ${\cal G}_{\overline{0}}$ and $t_1={1 \over 4}$, $t_2=1$, $t_3=-{1 \over 2}$. The N=2 hierarchies thus considered are generalizations of the particular bosonic $sl(3)$ hierarchies. With similar arguments as previously, it can be shown that the phase space may be reduced to four odd superfields. For example, in the case $i=1$, the Lax operator ${\cal L}$ takes the form
\begin{equation}
{\cal L} = D + 
\left( 
\begin{array}{cccc}
 0 & \phi_1& \phi_2 & \lambda\\
 \bar\phi_{1} &0 &0 & 0\\
 \bar\phi_{2}&0 &0 &0 \\
 \lambda &0 &0 &0 
\end{array}
\right) .
\end{equation} 
The linear problem ${\cal L}\Psi =0$ then leads to the eigenvalue equation
\begin{equation}
L \psi_4=  ( \partial - \sum_{k=1}^{2}\phi_{i} D^{-1} \bar\phi_{i} D) \psi_4 = \lambda^2\psi_4
\end{equation}
where one recognizes the N=1 scalar Lax operator of the two components N=2 KP hierarchy. The generalization to $sl(n \vert 1)$ is a priori straigthforward.

\end{document}